\documentclass[aps,twocolumn,amsmath,amssymb,showpacs,prl]{revtex4}
\usepackage{epsf}
\usepackage{graphicx}

\begin{document}

\title{Anomalous dressing of Dirac fermions in the topological surface state of Bi$_2$Se$_3$, Bi$_2$Te$_3$, and Cu$_x$Bi$_2$Se$_3$}

\author{Takeshi Kondo}
\affiliation{ISSP, University of Tokyo, Kashiwa, Chiba 277-8581, Japan}

\author{Y.~Nakashima}
\affiliation{ISSP, University of Tokyo, Kashiwa, Chiba 277-8581, Japan}

\author{Y.~Ota}
\affiliation{ISSP, University of Tokyo, Kashiwa, Chiba 277-8581, Japan}

\author{Y.~Ishida} 
\affiliation{ISSP, University of Tokyo, Kashiwa, Chiba 277-8581, Japan}

\author{W.~Malaeb} 
\affiliation{ISSP, University of Tokyo, Kashiwa, Chiba 277-8581, Japan}

\author{K.~Okazaki}
\affiliation{ISSP, University of Tokyo, Kashiwa, Chiba 277-8581, Japan}

\author{S.~Shin} 
\affiliation{ISSP, University of Tokyo, Kashiwa, Chiba 277-8581, Japan}

\author{M.~Kriener} 
\affiliation{Institute of Scientific and Industrial Research, Osaka University, Osaka 567-0047, Japan}

\author{Satoshi Sasaki} 
\affiliation{Institute of Scientific and Industrial Research, Osaka University, Osaka 567-0047, Japan}

\author{Kouji Segawa} 
\affiliation{Institute of Scientific and Industrial Research, Osaka University, Osaka 567-0047, Japan}

\author{Yoichi Ando} 
\affiliation{Institute of Scientific and Industrial Research, Osaka University, Osaka 567-0047, Japan}

\date{\today}
\begin{abstract}

Quasiparticle dynamics on the topological surface state of Bi$_2$Se$_3$,
Bi$_2$Te$_3$, and superconducting Cu$_x$Bi$_2$Se$_3$ are studied by 7 eV
laser-based angle resolved photoemission spectroscopy. We find strong
mode-couplings in the Dirac-cone surface states at energies of $\sim3$
and $\sim$15--20 meV, which leads to an exceptionally large coupling
constant $\lambda $ of $\sim3$, which is one of the strongest ever
reported for any material. This result is compatible with the recent
observation of a strong Kohn anomaly in the surface phonon dispersion of
Bi$_2$Se$_3$, but it appears that the theoretically proposed
``spin-plasmon" excitations realized in helical metals are also playing
an important role. Intriguingly, the $\sim3$ meV mode coupling is found
to be enhanced in the superconducting state of Cu$_x$Bi$_2$Se$_3$.

\end{abstract}


\maketitle

Topological insulators (TIs) are a new class of materials with Dirac
fermions appearing on the surface \cite{Hasan}. The nature of Dirac
fermions has already been actively studied in the graphitic materials
\cite{GraphenReview}, and it has been elucidated that the Dirac band
dispersion is anomalously renormalized by such effects as
electron-phonon interaction, electron-hole pair generation, and
electron-plasmon coupling, leading to various intriguing properties
\cite{QuasiGraphen,GraphenSTM,ChiangPRL}. While topological insulators
are essentially understood within the noninteracting topological theory
\cite{TI_PRL,TI_PRB}, the Dirac fermions realized in real materials
would be affected by nontrivial many body interactions, and hence the
investigation of the quasiparticle dynamics is important for extending
our understanding beyond the noninteracting regime. Since the Dirac
fermions in TIs are distinct form of those in graphitic materials in terms
of their helical spin texture as well as possible interactions with a
separate bulk electronic state, the low-energy excitations in the
topological surface state are of particular interest. Indeed, such
excitations are important not only for understanding many body
interactions and couplings to other degrees of freedom in the
topological surface sate, but also for assessing the stability of
putative Majorana fermions that are expected to emerge on the surface of
superconducting TIs \cite{TI_super1,TI_super2}.
 
In this context, there are already indications of the significance of
many body interactions in the topological surface state. For example, a
pronounced Kohn anomaly to indicate a strong electron-phonon coupling
was recently observed in the surface phonon branch of Bi$_2$Se$_3$
\cite{Kohn}; scanning tunneling spectroscopy (STS) uncovered an
intriguing feature with finely-resolved sharp peaks at low energies
($<20$ meV) in the Landau-level spectra \cite{Hanaguri}, pointing to an
anomalous increase in the quasiparticle lifetime near the Fermi energy
($E_{\rm F}$). Theoretically, it has been proposed that a novel low-energy
collective mode called ``spin-plasmon" would emerge as a consequence of
the spin-momentum locking in the topological surface state
\cite{SpinPlasmon}. Therefore, it is important to elucidate how the
Dirac dispersion is renormalized close to $E_{\rm F}$. However, so far
the angle-resolved photoemission spectroscopy (ARPES) has not been able
to detect any significant renormalization in the Dirac dispersions in
TIs \cite{Valla}, possibly because of the lack of sufficient energy
resolutions.

In this Letter, we demonstrate that the Dirac dispersion of the
topological surface state is indeed anomalously renormalized, by using
state-of-the-art ARPES with a 7-eV laser photon source. Availability of
the ultra-high energy resolution ($\sim1$ meV) and the extremely low
temperature ($\sim1$ K) \cite{Okazaki} enabled us to detect low-energy
kinks in the dispersion at $\sim$3 and $\sim$15--20 meV, giving evidence
for hitherto-undetected mode couplings. The analysis of the kinks leads
to the estimate of the coupling constant $\lambda$ of as large as
$\sim3$, which is one of the largest reported for any material
\cite{McMillan,ARPESsurface}. Despite the existence of this boson-mode
coupling, we observed no overall band reconstruction down to the lowest
temperature, indicating that the topological surface state is protected
from density-wave formations, which is usually expected to occur with
such an extremely strong coupling with bosons
\cite{Peierls,Bi_CDW,Bi_SpinOrbit,Rotenberg_review,Warping}.

Single crystals of Bi$_2$Se$_3$ and Bi$_2$Te$_3$ were grown by melting
stoichiometric amounts of elemental shots in sealed evacuated quartz
glass tubes. Superconducting samples of Cu$_{0.24}$Bi$_2$Se$_3$ with
$T_{\rm c}$ of 3.5 K and a shielding fraction of $\sim30\%$ [see
Fig. 3(a)] were prepared by electrochemically intercalating Cu into the
pristine Bi$_2$Se$_3$ \cite{Ando1,Ando2,Ando3,Ando4}. ARPES measurements were
performed using a Scienta R4000 hemispherical analyzer with an
ultraviolet laser ($h\nu$ = 6.994 eV) at the Institute for Solid State
Physics (ISSP), University of Tokyo \cite{Kiss1,Kiss2}.

%%%%%%%%%%%%%%%%%%%%%%
\begin{figure}
\includegraphics[width=3.2in]{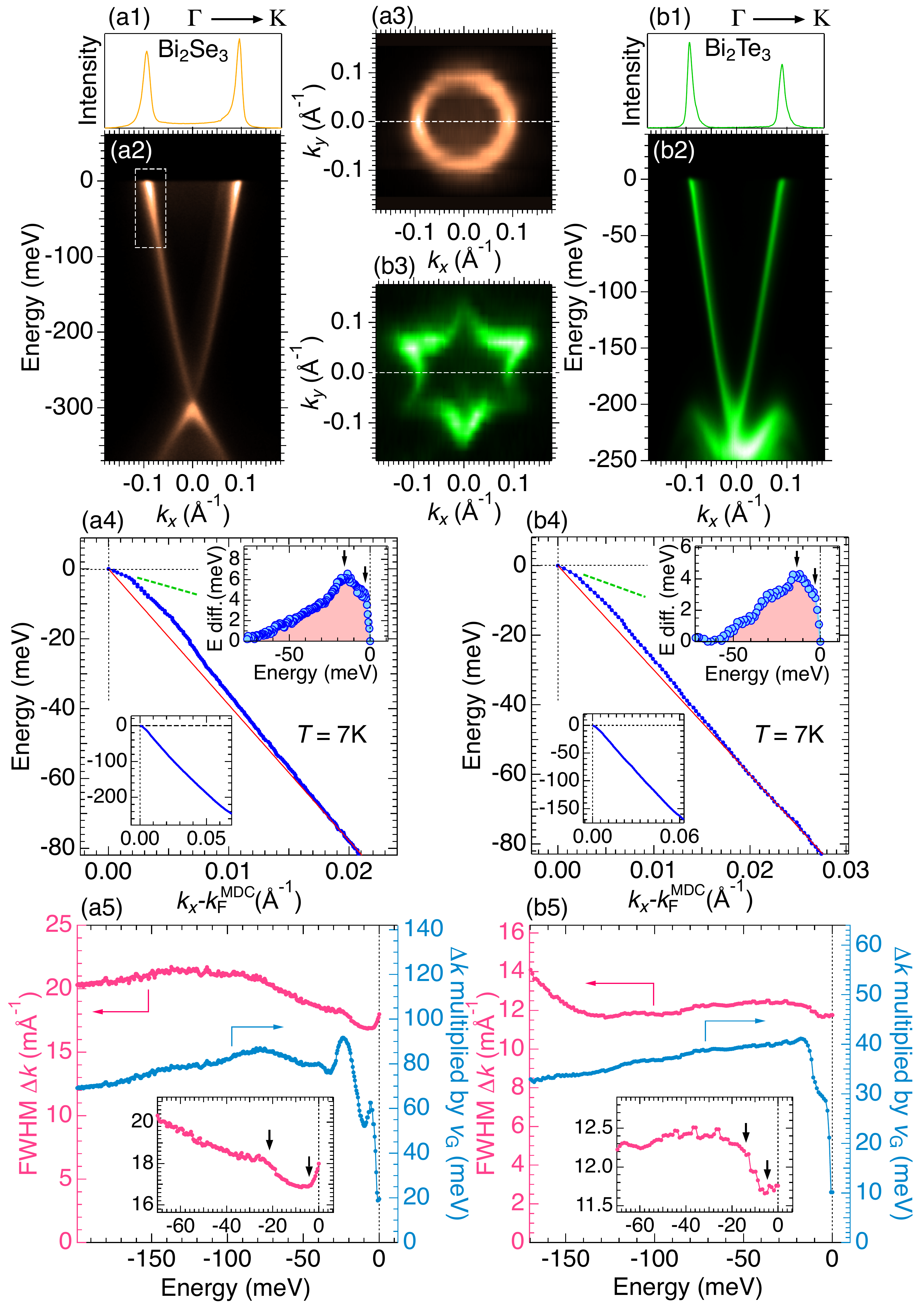}
\caption{(Color online)
Data for (a1)-(a5) Bi$_2$Se$_3$ and (b1)-(b5) Bi$_2$Te$_3$. (a1),
(b1) MDC at $E_{\rm F}$. (a2), (b2) Band dispersion map along $\Gamma$-K
[dashed lines in (a3) and (b3)]. (a3), (b3) Fermi surface map. (a4),
(b4) MDC-derived band dispersion. The same dispersion over
a wide energy range is shown in the lower-left inset. The upper-right
inset plots the energy difference from the linear dispersion.
(a5), (b5) MDC peak width $\Delta k$ (full width at
half maximum, FWHM), and the $\Delta k(E)$ multiplied by the
group velocity $v_G(E)$. The inset shows the
$\Delta k (E)$ close to $E_{\rm F}$.} 
\label{fig1}
\end{figure}
%%%%%%%%%%%%%%%%%%%%%%

%%%%%%%%%%%%%%%%%%%%%%
\begin{figure}
\includegraphics[width=3in]{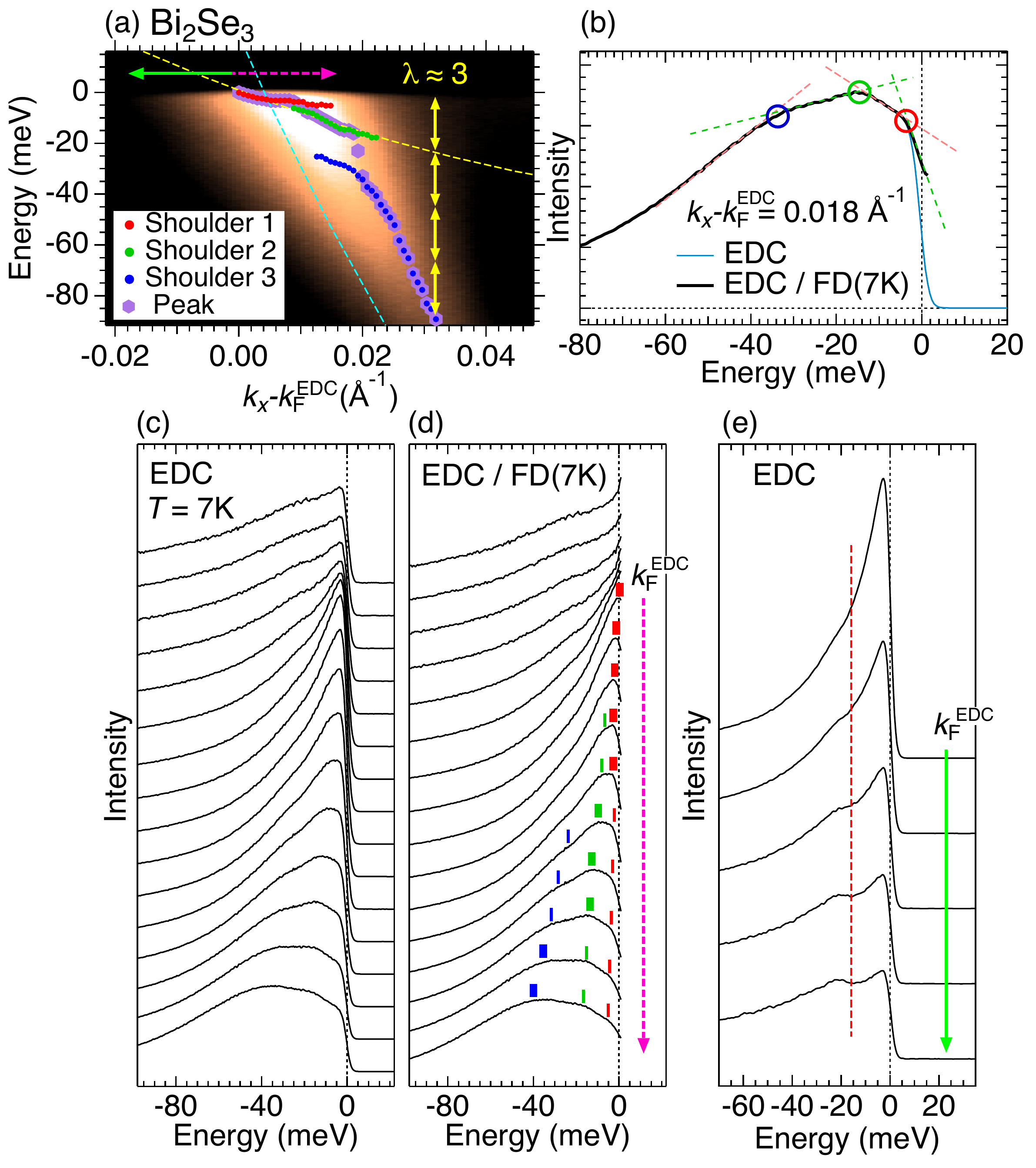}
\caption{(Color online)
Data for Bi$_2$Se$_3$ within a narrow range near $E_F$ marked in Fig. 1(a2).
(a) ARPES image; parabolic bands (dashed lines) with a mass of 0.83$m_e$
and 0.14$m_e$ \cite{Quantum} are superimposed. (b) Typical EDC with
three features. (c) EDCs near $k_F$ on the occupied-state side, and 
(d) those divided by Fermi function at
$T=7$ K. Energies
of shoulder-like structures [circles in (b) and bars in (d)] are plotted
on (a). The energy eigenvalue for each $k$, $\varepsilon (k)$, which is
determined by the energy positions at which the spectral intensity
becomes maximum, are indicated in (d) with bold bars, and plotted in (a)
with thick filled hexagons. (e) EDCs beyond $k_{\rm F}$ (unoccupied-state
side). The dashed line indicates the energy of the spectral dip.}
\label{fig1} 
\end{figure} 
%%%%%%%%%%%%%%%%%%%%%%

Figures 1(a1)-(a5) and 1(b1)-(b5) show the ARPES data of Bi$_2$Se$_3$
and Bi$_2$Te$_3$, respectively. The Dirac cones are clearly seen in the
dispersion maps [Figs. 1(a2) and 1(b2)], and the shapes of the Fermi
surface (FS) are very different between the two compounds [Figs. 1(a3)
and 1(b3)]. In the present experiment, we did not observe any
quantum-well states which emerge when adsorption of residual gases on
the sample surface causes charge doping
\cite{Aging_CoPb,Aging_Co,Aging_K,Aging_H2O,Hasan_perturbation,
2Dgas_NatComm,ImpurityScattering}. Also, our data are free from spectral
intensity from the bulk conduction band, as can be clearly seen in Figs.
1(a1) and 1(b1) where the momentum distribution curves (MDCs) at $E_{\rm
F}$ show only two sharp peaks from the Dirac dispersion. 

The novel feature in our data is that the MDC-derived band dispersions
[Figs. 1(a4) and 1(b4)] obviously deviate from straight lines, pointing
to a large mass enhancement; the renormalized slope of the dispersion
close to $E_F$ is shown by dashed lines. As shown in the upper insets of
Figs. 1(a4) and 1(b4), we calculate the energy difference between the
putative linear dispersion (which is expected when the mode couplings
are absent) and the measured one to estimate the strength of the
coupling as a function of energy. We found anomalies at two energies,
$\sim -15$ and $\sim -3$ meV, indicative of electron couplings with two
different kinds of collective modes. 

The effects of the couplings should also be observed in the energy
dependence of the MDC peak width ($\Delta k$) because of the
Kramers-Kronig relation between ${\mathop{\rm Re}\nolimits} \Sigma$ and
${\mathop{\rm Im}\nolimits} \Sigma$ ($\Sigma$ is the self-energy). In
Figs. 1(a5) and 1(b5), we plot the obtained spectrum of $\Delta k$ for
Bi$_2$Se$_3$ and Bi$_2$Te$_3$, respectively; as expected, $\Delta k(E)$
presents kinks at the two energy scales, $\sim -$(15-20) and $\sim -3$
meV, which are better seen in the insets. The $\sim -3$ meV kink marks
the onset of an anomalous increase in the magnitude of $\Delta k(E)$
toward $E_{\rm F}$. 

To understand the complex behavior of the band dispersions revealed at
low energy, we examine the shapes of the energy distribution curves
(EDCs) around $k_{\rm F}$ shown in Fig. 2(c). In Fig. 2(d), those
original EDCs are divided by the Fermi function at the measured
temperature of 7 K convoluted with the experimental energy resolution,
to remove the effect of Fermi cut-off. In the resulting curves, one can
identify up to three shoulder-like features [an example for $k_x -
k_F^{\rm EDC} = 0.018 {\rm \AA}^{-1}$ is shown in Fig. 2(b)], and the
energy positions of those features are plotted on the ARPES image shown
in Fig. 2(a). One can see in Fig. 2(d) that one of the shoulder-like
features on the curves actually corresponds to the maximum; the
dispersion of this maximum close to $k_F$ is also plotted in Fig. 2(a)
with thick filled symbols, and this dispersion is fitted with a
parabolic dashed line in Fig. 2(a), giving a significantly enhanced
effective mass of 0.83$m_e$ ($m_e$ is the free-electron mass). For
comparison, we also plot a putative band dispersion with a mass of
0.14$m_e$, which was estimated for the bulk band from quantum
oscillations \cite{Quantum}. The mass enhancement realized in the
topological state is remarkable. Even more surprisingly, the estimated
value of $\lambda = {v_0}/{v_{\rm F}} - 1$ ($v_{\rm 0}$ and $v_{\rm F}$
are the bare-electron velocity \cite{F_note} and the renormalized Fermi velocity, 
respectively) is as large as $\sim$ 3 as demonstrated in Fig.2(a),
which is one of the strongest coupling ever reported in any material
\cite{McMillan,ARPESsurface}.

%%%%%%%%%%%%%%%%%%%%%%
\begin{figure}
\includegraphics[width=3.2in]{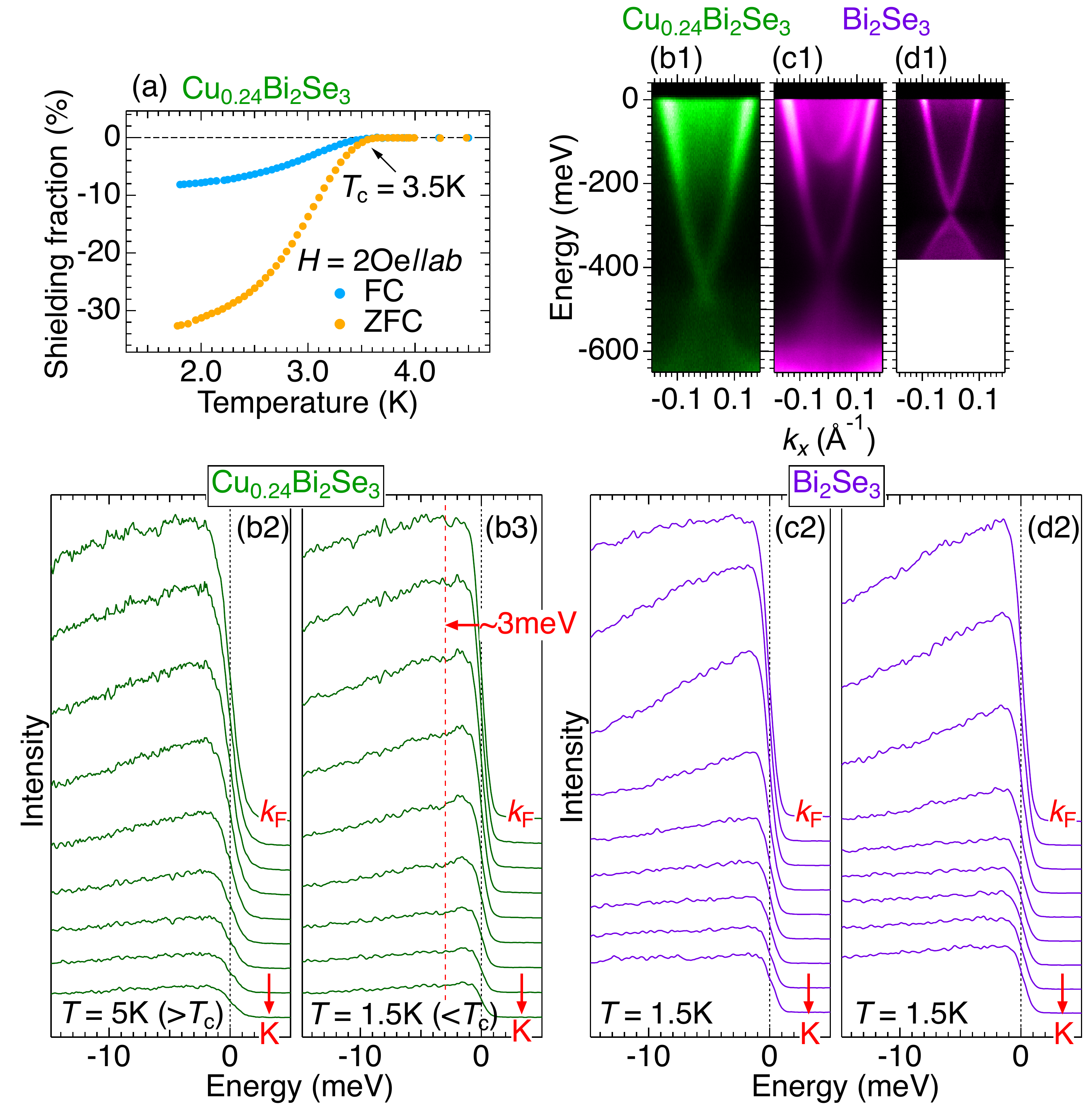}
\caption{(Color online)
Data for Cu$_{0.24}$Bi$_2$Se$_3$ superconductor with $T_{\rm c}$ = 3.5 K
and the pristine Bi$_2$Se$_3$. (a) Field-cooled (FC) and zero-field-cooled (ZFC) 
data of the superconducting shielding fraction of the sample used for 
ARPES experiments. Band dispersion map along $\Gamma$-K for (b1)
Cu$_{0.24}$Bi$_2$Se$_3$, (c1) aged and (d1) fresh surfaces of
Bi$_2$Se$_3$. EDCs of
Cu$_{0.24}$Bi$_2$Se$_3$ close to $k_{\rm F}$ measured (b2) above $T_{\rm c}$
 and (b3) below $T_{\rm c}$. The dashed line in (b3) indicates the
spectral dip. EDCs of the pristine Bi$_2$Se$_3$ close to $k_{\rm F}$ for
(c2) aged and (d2) fresh surfaces. } 
\label{fig1}
\end{figure} 
%%%%%%%%%%%%%%%%%%%%%%

%%%%%%%%%%%%%%%%%%%%%%
\begin{figure}
\includegraphics[width=3in]{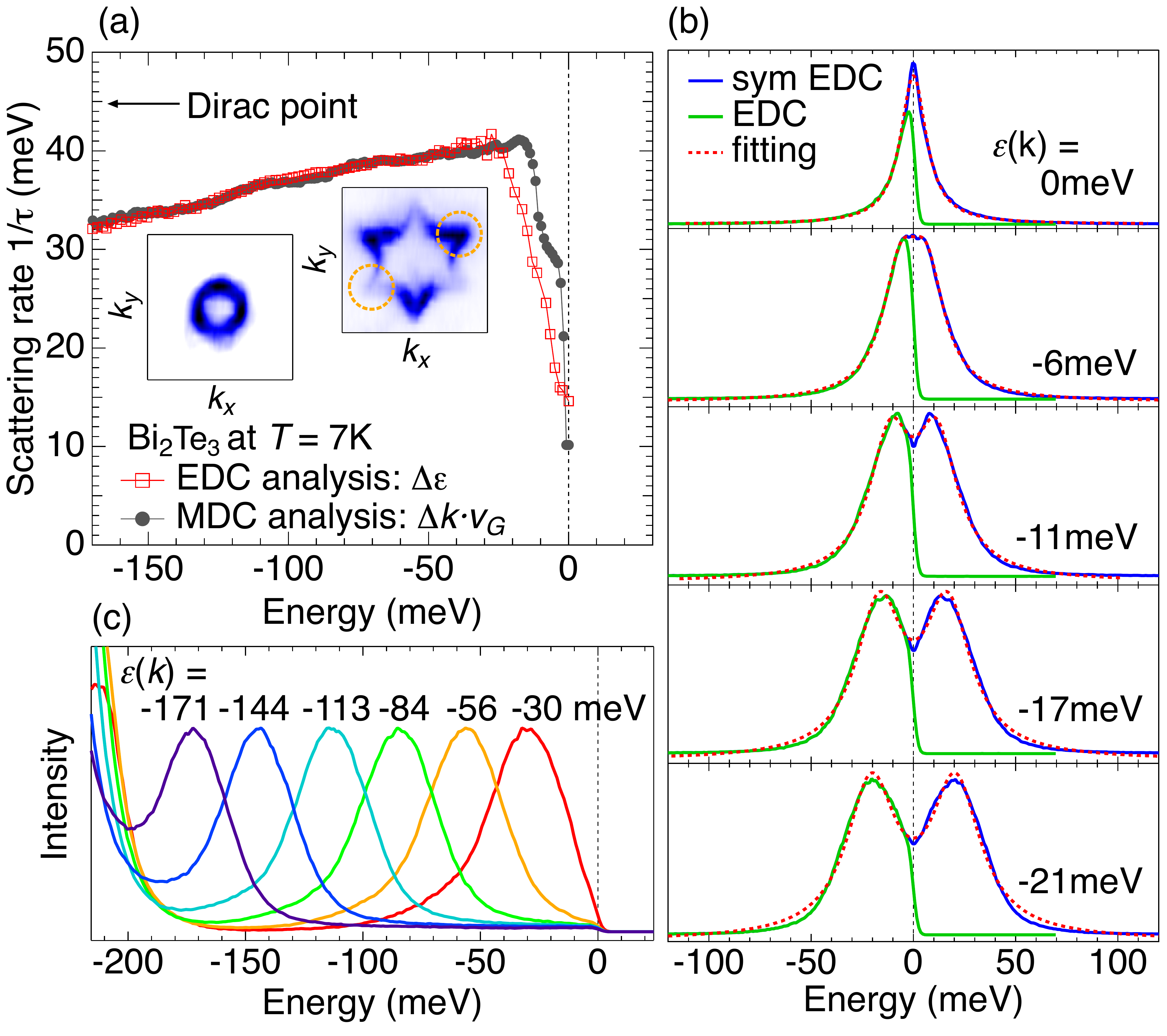}
\caption{(Color online)
Results for Bi$_2$Te$_3$. (a) Scattering rate ($1/\tau$) estimated
from EDC peak width $\Delta \varepsilon$ (full width at half maximum, 
FWHM) as well as from the MDC peak width multiplied by the group velocity, $\Delta k
\cdot {v_{\rm G}}$. Insets show the ARPES map at -150 meV (left) and
$E_{\rm F}$ (right). (b) Estimation of $\Delta \varepsilon$ by fitting
double Lorentzians to the symmetrized EDCs. (c) EDCs far from $k_{\rm
F}$; the energy eigenvalue for each $k$, $\varepsilon (k)$, corresponds
to the peak position in each EDC. 
} 
\label{fig1}
\end{figure}
%%%%%%%%%%%%%%%%%%%%%%

Perhaps the most direct way to demonstrate the strong mode-coupling is to
present the peak-dip-hump structure in the EDCs. In Fig. 2(e), we show
EDCs at $k_{\rm F}$ and beyond, where the peak-dip-hump shape is usually
emphasized, and indeed, a clear dip can be seen at $\sim -$16 meV
(dashed line). However, a similar structure associated with the $\sim 3$
meV mode is not clearly visible in the data of Bi$_2$Se$_3$. 

Intriguingly, we found that a peak-dip-hump structure coming from the
$\sim 3$ meV mode coupling becomes visible in the superconducting
samples of Cu-doped Bi$_2$Se$_3$. Figures 3(b2) and 3(b3) shows the EDCs
of Cu$_{0.24}$Bi$_2$Se$_3$ with $T_{\rm c}$ = 3.5 K measured above
and below $T_{\rm c}$, respectively. The peak-dip-hump structure is
seen at $\sim 3$ meV below $T_{\rm c}$ but 
it is gone above $T_{\rm c}$. We also measured the
pristine Bi$_2$Se$_3$ at the same condition ($T$ = 1.5 K), but did not
observe the peak-dip-hump [Fig. 3(d2)]. 
Furthermore, we doped the surface of the
pristine sample up to a doping level similar to that of
Cu$_{0.24}$Bi$_2$Se$_3$ by exposing it to residual gases 
[compare Figs. 3(c1) and (b1)], but again, the peak-dip-hump structure is
not observed [Fig. 3(c2)]. Obviously, the enhancement of the $\sim 3$ 
meV mode coupling has something to do with the superconductivity, and 
the origin of this enhancement needs to be scrutinized in future studies. 
In passing, we note that we did not
detect any signature of the superconducting energy gap in the present
experiment, probably because of the nonuniform surface condition after
the sample cleaving. More elaborate studies would be required to nail
down the topological superconductivity in Cu$_x$Bi$_2$Se$_3$ \cite{Ando3}
by ARPES experiments.

Now we show the relevance of the mode couplings to the scattering rate
($1/\tau$). Since the MDC width ($\Delta k$) is not exactly the same as
$1/\tau$ for a curved band-dispersion, we have extracted the scattering
rate with two methods: one simply uses the peak width of the EDCs
($\Delta\varepsilon = 1/\tau$), the raw data of which are shown in Figs.
4(b) and 4(c); the other calculates the product of the MDC width and the
experimentally obtained group velocity ($\Delta k \cdot {v_{\rm G}}
\approx 1/\tau$). The behavior of $\Delta k \cdot {v_{\rm G}}$ in
Bi$_2$Se$_3$ and Bi$_2$Te$_3$ are plotted in Figs. 1(a5) and 1(b5),
respectively, which clearly reflect the mode couplings at $\sim 20$ and
$\sim 3$ meV. Intriguingly, as shown in Fig. 4(a) for Bi$_2$Te$_3$, both
$\Delta\varepsilon$ and $\Delta k \cdot {v_{\rm G}}$ present a sharp
reduction toward $E_{\rm F}$. It is useful to note that this behavior is
consistent with the STS result showing a sharpening of Landau-level
peaks at low energies ($<20$ meV) \cite{Hanaguri}. 

One may also notice in Fig. 4(a) that $1/\tau$ gradually decreases
toward the Dirac point. This is unusual, because a monotonic {\it
increase} in the electron-electron interaction with increasing binding
energy is usually expected in conventional metals \cite{Valla_Mo}. We
speculate that this unusual behavior is a consequence of the fact that
the penetration depth of the surface state increases as the momentum
moves away from the Dirac point \cite{BandCal_NewJour}, which makes the
surface state to gradually gain some bulk character. In fact, a feature
to suggest such a variation is seen in the energy-contour maps plotted
in the inset of Fig. 4(a): the contour near the Dirac point (left image)
is almost circular, but close to $E_{\rm F}$ (right image), it exhibits
a $C_{3}$ modulation which reflects the symmetry of the bulk. 

Now we discuss the most crucial question, namely, the origin of the
bosons causing the anomalies at $\sim -$(15-20) and $\sim- 3$ meV in the
ARPES spectra. A plausible candidate for the higher binding-energy
one is the out-of-plane optical phonon mode ${\rm{A}}_{1g}^2$
with $\omega = 21$ and 16 meV for Bi$_2$Se$_3$ and Bi$_2$Te$_3$,
respectively \cite{RamanOld,RamanNew}. It seems that $\Delta k$ begins
to decrease toward $E_{\rm F}$ [Figs. 1(a5) and 1(b5)] at almost the
same energy as that of the ${\rm{A}}_{1g}^2$ mode. Also, the relevance
of the phonon coupling is supported by the fact that the mode energy
observed in Bi$_2$Se$_3$ ($\sim 20$ meV) is higher than that in
Bi$_2$Te$_3$ ($\sim 15$ meV), which is consistent with the mass
difference between Se and Te ($\sqrt
{{m_{{\rm{Te}}}}/{m_{{\rm{Se}}}}}=1.27$).

As for the $\sim3$ meV mode, there are two possible origins. One is the
optical mode of surface phonons. Recently, a strong Kohn anomaly was
detected by a helium atom surface scattering (HASS) experiment in a
phonon branch of Bi$_2$Se$_3$ \cite{Kohn} at approximately 2$k_{\rm
F}^{\rm Dirac}$ ($k_{\rm F}^{\rm Dirac}$ is the Fermi momentum on the
Dirac cone) with the characteristic energy of $\sim$3 meV, and this Kohn
anomaly was attributed to the surface optical phonon mode \cite{Kohn}.
However, Kohn anomaly is typically accompanied by an extensive FS
nesting \cite{Kohn_W,Rotenberg_W,Rotenberg_review}, which is absent in
Bi$_2$Se$_3$ [see Fig.1(a3)]. Also, the anomaly observed in the HASS
experiment was isotropic \cite{Kohn}, while the Kohn anomaly is usually
very anisotropic \cite{Kohn_W,Rotenberg_W,Rotenberg_review}. Hence,
the Kohn anomaly in Bi$_2$Se$_3$ is quite unusual. In our ARPES data,
the $\sim3$ meV mode coupling in Bi$_2$Te$_3$ is obviously weaker than
that in Bi$_2$Se$_3$, even though in Bi$_2$Te$_3$ the FS nesting on the
Dirac cone is more pronounced \cite{Warping} and hence the Kohn anomaly
is expected to be stronger. In addition, the $\lambda$ value of 0.43 has
been obtained in the HASS experiment on Bi$_2$Se$_3$ for the relevant
phonons \cite{Kohn_self}, but this value is obviously too small to
account for the very strong coupling observed here for the $\sim 3$ meV
mode. Therefore, the surface optical phonons alone are obviously not
sufficient for understanding the lower energy mode, and we need to seek
for additional ingredients. 

In this respect, another, more promising, origin of the $\sim 3$ meV
mode is the theoretically proposed ``spin-plasmon", which is suggested
to have a maximum energy of $\sim2.2$ meV \cite{SpinPlasmon}. This mode
consists of coupled plasmons and spin waves, and unlike the Kohn
anomaly, it is expected for the round Fermi surface as in Bi$_2$Se$_3$
\cite{SpinPlasmon}. The observed upturn in the MDC width toward $E_F$ [see insets of
Figs. 1(a5) and 1(b5)], which is not expected in a typical
electron-phonon coupling, can be interpreted to signify an increasingly
stronger interactions of the Dirac quasiparticles with spin-plasmons 
near $E_{\rm F}$. Note that such strong interactions
between the two are expected only when the plasmon spectrum does not
overlap with the continuum of electron-hole excitations
\cite{QuasiGraphen,LEGPlasmon}, and hence the plasmon coupling should
dominate the scatterings with $q \sim 0$. Therefore, it is natural that
the quasiparticle scattering is enhanced toward $E_{\rm F}$ in this
spin-plasmon scenario. All told, it is most likely that the large angle
scattering ($q \sim 2k_{\rm F}$) by the surface optical phonons and the
small angle scattering ($q \ll 2k_{\rm F}$) by the spin-plasmons both
are playing roles in the enormously strong mass enhancement observed
near $E_{\rm F}$ on the topological surface state.

In conclusion, we have investigated the quasiparticle dynamics in the
topological surface state of Bi$_2$Se$_3$, Bi$_2$Te$_3$, and
Cu$_x$Bi$_2$Se$_3$. We found strong mode-couplings at the binding energy
of $\sim$15--20 and $\sim3$ meV. The coupling to the ${\rm{A}}_{1g}^2$
phonons is proposed as the candidate for the former mode. As for the
$\sim3$ meV mode, there are two possible origins. One is the optical
mode of surface phonons. The other is the spin-plasmons, which are
theoretically proposed as low-energy excitations of the helically
spin-polarized Dirac fermions. Intriguingly, despite the extremely large
mass enhancement factor $\lambda$ of $\sim3$, the topological surface
state remains free from any band reconstruction down to the lowest
temperature, indicating that the helical Dirac cone is protected from
density-wave formations which are naturally expected for a system with
extremely strong couplings to bosons.

This work is supported by JSPS (FIRST Program, NEXT Program, and KAKENHI 24740218), MEXT
(Innovative Area ``Topological Quantum Phenomena" KAKENHI 22103004), and
AFOSR (AOARD 124038).


\begin{thebibliography}{99}

\bibitem{Hasan}
M. Z. Hasan and C. L.  Kane, 
Rev. Mod. Phys. {\bf 82}, 3045 (2010).

\bibitem{GraphenReview}
A. H. Castro Neto $et\ al$., 
Rev. Mod. Phys. {\bf 81}, 109 (2009).

\bibitem{QuasiGraphen}
A. Bostwick$et\ al$., 
Nature Phys. {\bf 3}, 36 (2007).

\bibitem{GraphenSTM}
Y. Zhang $et\ al$., 
Nature Phys. {\bf 4}, 627 (2008).

\bibitem{ChiangPRL}
Y. Liu $et\ al$., 
Phys. Rev. Lett.  {\bf 105}, 136804 (2010).

\bibitem{TI_PRL}
L. Fu, C. L. Kane, and E. J. Mele, 
Phys. Rev. Lett.  {\bf 98}, 106803 (2007).

\bibitem{TI_PRB}
J. E. Moore and L. Balents, 
Phys. Rev. B {\bf 75}, 121306(R) (2007).

\bibitem{TI_super1}
L. Fu and C.L. Kane, 
Phys. Rev. Lett.  {\bf 100}, 096407 (2008).

\bibitem{TI_super2}
X. -L. Qi, T. L. Hughes, S. Raghu, and S. -C. Zhang , 
Phys. Rev. Lett.  {\bf 102}, 187001 (2009).

\bibitem{Kohn}
X. Zhu $et\ al$., 
Phys. Rev. Lett.  {\bf 107}, 186102 (2011).

\bibitem{Hanaguri}
T. Hanaguri, K. Igarashi, M. Kawamura, H. Takagi, and T. Sasagawa, 
Phys. Rev. B {\bf 82}, 081305(R) (2010).

\bibitem{SpinPlasmon}
S. Raghu, S. B. Chung, X. -L. Qi, and S. -C. Zhang, 
Phys. Rev. Lett.  {\bf 104}, 116401 (2010).

\bibitem{Valla}
Z.-H Pan $et\ al$., 
Phys. Rev. Lett.  {\bf 108}, 187001 (2012).

\bibitem{Okazaki}
K. Okazaki $et\ al$., 
Science {\bf 337}, 1314 (2012).

\bibitem{McMillan}
W. L. McMillan, 
Phys. Rev. {\bf 167}, 331 (1968).

\bibitem{ARPESsurface}
P. Hofmann, I. Y. Sklyadneva, E. D. L. Rienks and E. V. Chulkov, 
New J. Phys. {\bf 11}, 125005 (2009).

\bibitem{Peierls}
K. Nasu, 
Phys. Rev. B {\bf 44}, 7625 (1991).

\bibitem{Bi_CDW}
C. R. Ast and H. H\"ochst, 
Phys. Rev. Lett.  {\bf 90}, 016403 (2003).

\bibitem{Bi_SpinOrbit}
Y. M. Koroteev $et\ al$., 
Phys. Rev. Lett.  {\bf 93}, 046403 (2004).

\bibitem{Rotenberg_review}
E. W. Plummer $et\ al$., 
Prog. Surf. Sci. {\bf 74}, 251 (2003).

\bibitem{Warping}
Liang Fu, 
Phys. Rev. Lett.  {\bf 103}, 266801 (2009).

\bibitem{Ando1}
M. Kriener $et\ al$., 
Phys. Rev. B {\bf 84}, 054513 (2011).

\bibitem{Ando2}
M. Kriener, K. Segawa, Z. Ren, S. Sasaki, and Y. Ando,
Phys. Rev.  Lett. {\bf 106}, 127004 (2011).

\bibitem{Ando3}
S. Sasaki $et\ al$., 
Phys. Rev.  Lett. {\bf 107}, 217001 (2011).

\bibitem{Ando4}
M. Kriener, K. Segawa, S. Sasaki, and Y. Ando, 
Phys. Rev.  B {\bf 86}, 180505(R) (2012).

\bibitem{Kiss1}
T. Kiss $et\ al$.,
Phys. Rev. Lett.  {\bf 94}, 057001 (2005).

\bibitem{Kiss2}
T. Kiss $et\ al$.,  
Rev. Sci. Instrum. {\bf 79}, 023106 (2008).

\bibitem{Aging_CoPb}
P. D. C. King $et\ al$., 
Phys. Rev. Lett.  {\bf 107}, 096802 (2011).

\bibitem{Aging_Co}
M. Bianchi, R. C. Hatch, J. Mi, B. B. Iversen, and P. Hofmann, 
Phys. Rev. Lett.  {\bf 107}, 086802 (2011).

\bibitem{Aging_K}
Z.-H. Zhu $et\ al$., 
Phys. Rev. Lett.  {\bf 107}, 186405 (2011).

\bibitem{Aging_H2O}
H. M. Benia, C. Lin, K. Kern, and C. R. Ast, 
Phys. Rev. Lett.  {\bf 107}, 177602 (2011).

\bibitem{Hasan_perturbation}
L. A. Wray $et\ al$., 
Nature Phys. {\bf 7}, 32 (2011).

\bibitem{2Dgas_NatComm}
M. Bianchi $et\ al$., 
Nature Commun. {\bf 1}, 128 (2010.)

\bibitem{ImpurityScattering}
S. R. Park $et\ al$., 
Phys. Rev. B {\bf 81}, 041405(R) (2010).

\bibitem{Quantum}
K. Eto, Z. Ren, A. A. Taskin, K. Segawa, and Y. Ando, Phys. Rev.  B {\bf 81}, 195309 (2010).

\bibitem{F_note} 
$v_0$ was estimated from the
slope of the putative linear dispersion in the absence of
mode-couplings, shown as solid straight lines in Figs. 1(a4) and 1(b4).

\bibitem{Valla_Mo}
T. Valla, A. V. Fedorov, P. D. Johnson, and S. L. Hulbert, 
Phys. Rev. Lett. {\bf 83}, 2085 (1999).

\bibitem{BandCal_NewJour}
W. Zhang, R. Yu, H.-J. Zhang, X. Dai, and Z. Fang, 
New J. Phys. {\bf 12}, 065013 (2010).

\bibitem{RamanOld}
W. Richter, H. Kohler, and C. R. Becker,
Phys. Status Solidi B {\bf 84}, 619 (1977).

\bibitem{RamanNew}
K. M. F. Shahil, M. Z. Hossain, D. Teweldebrhan, and A. A. Balandin,
Appl. Phys. Lett. {\bf 96}, 153103 (2010).

\bibitem{Kohn_W}
E. Hulpke and J. L\"udecke, 
Phys. Rev.  Lett. {\bf 68}, 2846 (1992).

\bibitem{Rotenberg_W}
E. Rotenberg, J. Schaefer, and S. D. Kevan,
Phys. Rev.  Lett. {\bf 84}, 2925 (2000).

\bibitem{Kohn_self}
X. Zhu $et\ al$., 
Phys. Rev. Lett.  {\bf 108}, 185501 (2012).

\bibitem{LEGPlasmon}
P. Hawrylak, 
Phys. Rev. Lett.  {\bf 59}, 485 (1987).

\end{thebibliography}
\end{document}